# All-optical spiking neurosynaptic networks with self-learning capabilities

J. Feldmann[1], N. Youngblood[2], C.D. Wright[3], H. Bhaskaran[2] and W.H.P. Pernice[1]*

[1]Institute of Physics, University of Muenster, Heisenbergstr. 11, 48149 Muenster, Germany
[2]Department of Materials, University of Oxford, Parks Road, OX1 3PH Oxford, UK
[3]Department of Engineering, University of Exeter, Exeter, EX4 QF, UK
*Correspondence to: wolfram.pernice@uni-muenster.de.

**Software-implementation, via neural networks, of brain-inspired computing approaches underlie many important modern-day computational tasks, from image processing to speech recognition, artificial intelligence and deep learning applications. Yet, differing from real neural tissue, traditional computing architectures physically separate the core computing functions of memory and processing, making fast, efficient and low-energy brain-like computing difficult to achieve. To overcome such limitations, an attractive and alternative goal is to design direct hardware mimics of brain neurons and synapses which, when connected in appropriate networks (or neuromorphic systems), process information in a way more fundamentally analogous to that of real brains. Here we present an all-optical approach to achieving such a goal. Specifically, we demonstrate an all-optical spiking neuron device and connect it, via an integrated photonics network, to photonic synapses to deliver a small-scale all-optical neurosynaptic system capable of supervised and unsupervised learning. Moreover, we exploit wavelength division multiplexing techniques to implement a scalable circuit architecture for photonic neural networks, successfully demonstrating pattern recognition directly in the optical domain using a photonic system comprising 140 elements. Such optical implementations of neurosynaptic networks promise access to the high speed and bandwidth inherent to optical systems, which would be very attractive for the direct processing of telecommunication and visual data in the optical domain.**

In our everyday life, artificial neural networks (ANNs) are already heavily active behind the scenes, for example carrying out tasks such as face and speech recognition that are frequently performed on our mobile phones[1]. Thinking of more complex applications, such as medical diagnostics[2] and autonomous driving[1][3], high-speed data-analysis will become even more important in the future. However, fulfilling this demand for fast and efficient processing using traditional processing and computation techniques is problematic, due to speed and energy inefficiencies[4]. Traditional computers are built following the von-Neumann architecture, having two separate units for memory and processor and operating in a sequential way one command at a time. Compared to the massively parallel signal processing of the brain, it becomes clear why simulating a neural network in software on a machine based on the von-Neumann architecture and limited by the transfer of data between memory and processor, cannot be efficient[5]. A more radical approach – neuromorphic computing – seeks to overcome the limitations of carrying out brain-like processing using conventional computers by developing hardware mimics of the basic building blocks of biological brains, i.e. neurons and synapses, and combining these into suitably-scaled networks and arrays. Such an approach could, for example, enable the efficient processing and analysis of data in parallel directly on-chip, so finding widespread utility in power-critical situations such as for mobile devices and so-called "edge computing" applications[6].

Recently, a number of different concepts for realizing hardware (i.e. neuromorphic) implementations for artificial intelligence (AI) have been proposed in the electrical domain[7][8] but optical approaches are very much in their infancy [9–13] A most promising candidate for photonic neuromorphic computing is however that based around phase-change materials and devices, since these have been shown to exhibit an intrinsic ability to provide in hardware the basic integrate-and-fire functionality of neurons and the plastic weighting operation of synapses[14][15][16][17][18]. However, a fully optical, integrated and scalable neuromorphic framework for implementing spiking neural networks using phase-change materials has—to the best of our knowledge—not yet been demonstrated. In this work, therefore, we propose and fabricate an all-optical spiking neuron circuit, with integrated all-optical synapses, and demonstrate that such a system is capable of the prototypical AI task of pattern recognition. Moreover, training/learning in the system is implemented in both a supervised and an unsupervised way, both cases having a wide range of applications but relying on different learning rules. In the first case, where training sets with pairs of known inputs and outputs are present, supervised learning rules can be applied, such as the well-known error backpropagation algorithm[19][20]. The second case requires unsupervised learning, meaning that the network adapts on its own to specific repeating features and patterns that are unknown beforehand[21]. We devise a layered architecture, based on wavelength division multiplexing (WDM), for realizing such complex integrated photonic systems, presenting a photonic neural network consisting of four neurons and sixty synapses (and 140 optical elements in total) which is able to successfully recognize letters presented to it. By implementing an all-optical spiking neural network on a nanophotonic chip, we provide a first step towards optical neuromorphic systems, which benefit from the high bandwidth and fast signaling properties that come with operating fully in the optical domain [22][23].

A sketch of our optical spiking neuron circuit, also incorporating all-optical synapses, and how it is integrated on a nanophotonic platform is shown in Figure 1a-b. The neuron represents a system comprising $N$ input (pre-synaptic) neurons, one output (post-synaptic) neuron and $N$ interconnecting synapses. Each connection between the pre-synaptic neurons and the post-synaptic neuron has a certain weight $w_i$. In the configuration of Figure 1a, optical pulses from the pre-synaptic neurons are fed from the left into the connecting synapses, thence to the post-synaptic neuron itself. The synapses are built of optical waveguides (see Figure 1b) and weighting is achieved via phase-change material cells (here of area 3.6 μm$^2$ and shown as red squares) integrated on top of the waveguides, which can modify the propagating optical mode in a controlled manner. Phase-change materials (PCMs) are commonly used in re-writable optical disc technologies, such as Blu-ray RE, and exhibit a large contrast in the absorption of light between their amorphous and crystalline states (phases) of matter[24][25][26]. With the PCM-cell in the amorphous state, the synaptic waveguide is highly transmissive, representing a strong connection between two neurons. In the crystalline state, however, most of the light is absorbed leading to a weak connection. After the input pulses have been weighted, they are combined into a single waveguide using a wavelength division multiplexing (WDM) [27,28] and guided to the (output) spiking neuron circuit. This is composed of a ring resonator with its own integrated PCM-cell that can be switched (between crystalline and amorphous states) by the incoming combined pulses. Switching the neuronal PCM-cell in turn changes the optical resonance condition of the ring and its propagation loss. When the neuronal PCM-cell is in the crystalline state, a suitable probe pulse sent along the 'output' waveguide couples strongly into the ring resonator and so no output pulse (spike) will be observed. However, if the combined power of the weighted input pulses from the pre-synaptic neurons is high enough to switch the neuronal PCM-cell to its amorphous state, the probe pulse is no longer on resonance with the ring and will be transmitted past the ring, so generating an output neural spike. As the switching of the PCM-cell only occurs above a certain threshold power, the neuron only generates an output pulse (spike) if the weighted sum of the input power exceeds this threshold. Thus, the system naturally emulates the basic integrate-and-fire functionality of a biological neuron. This artificial neuron, shown in photonic circuit form in Figure 1c, serves as a building block in layered photonic networks (described later) suited to the scalable implementation of neurosynaptic systems.

The neurosynaptic system described so far in the above provides the basic structure needed for supervised learning tasks, where the weights of the inputs are set by an external supervisor.

However, in order to also be capable of unsupervised learning, we add a feedback waveguide channelling parts of the neuron's output spike back to the synaptic PCM-cells. In this way (and as described in more detail later), the connections from all inputs that contributed to a particular output spike will be enhanced, while those that did not contribute will be weakened – or in Hebbian terms, "neurons that fire together, wire together."

Figure 1d shows an optical micrograph of the actual implementation (x3) of a single-neuron neurosynaptic system, fabricated via electron-beam lithography on a silicon-nitride on silicon-oxide platform. Several input waveguides each with a synaptic PCM-cell on top (red circles) are fed to the upper waveguide using small ring resonators as a simple multiplexing device (the ring resonators have four different radii increasing linearly from 40 µm to 55 µm, an optical Q of around 10.000 and provide insertion loss of 1.5 dB - see supplementary section S5 for more details). This upper waveguide then leads the light to the neuronal (large) ring resonator (radius 60 µm) with its own integrated PCM-cell (area 9 µm$^2$). Probe pulses sent to the waveguide lying below the neuronal ring resonator either couple to it or generate an output spike depending, as described previously, on whether the neuronal PCM-cell is in the crystalline or amorphous state. The light is coupled onto and off the chip using grating couplers which provide access to multiple optical fibres in the measurement setup.

Figure 2a shows in detail the photonic signal processing and optical operation of a single neuron. Each input pulse is sent on a different wavelength $\lambda_i$ and firstly partly absorbed (weighted) by the relevant synaptic PCM-cell. After weighting, the individual input waveguides are combined with a multiplexer to a single waveguide, summing up the input powers (more details on the multiplexing are given in the supplementary information). If this power is high enough to switch the neuronal PCM-cell of the large ring resonator (see Figure 2b), an output spike is generated and in the case of unsupervised learning the synaptic weights are adjusted using the feedback loop. In this specific device/example an output pulse is generated if the switching power exceeds 430 pJ (see section 9 in the supplementary materials). In Figure 2b a more detailed scanning electron micrograph of the ring resonator used to deliver the spiking neuron function can be seen. The neuronal PCM-cell used to tune the resonance condition is deposited on top of a waveguide crossing specially designed for low optical losses (0.23 dB [29]). The second waveguide crossing (without a PCM-cell) is only used for testing purposes and offers the ability of a bias input.

In order to characterise the integrate-and-fire type response (or activation function) of our neuron circuit, several transmission spectra for the resonator were obtained after sending pulses of different energy to the neuronal PCM-cell via the crossing waveguide (Figure 2c). The transmission spectrum undergoes a significant change between the initial crystalline state (0% pulse energy) and the final amorphous state (maximum pulse energy). This is partially explained by the resonance wavelength (at which the light is coupled into the ring and therefore the transmission is minimal) slightly shifting to shorter wavelength with increasing amount of amorphization (which causes a change in the real part of the refractive index of the phase-change material, leading to a slightly shifted resonance condition (see Supplementary Materials)). However, we also observe that the minimum transmission after amorphizing the PCM-cell is much higher. This is a combined effect of the change in the imaginary part of the refractive index (absorption) and a change in the extinction ratio of the resonator. The lower absorption in the amorphous phase obviously leads to higher transmission but, equally important, changing the loss per round trip in the ring affects the coupling between ring and waveguide and therefore alters the extinction ratio. By plotting the transmission of the ring resonator after different excitation pulses vs. the pulse energy at a fixed wavelength, an activation function as shown in Figure 2d is obtained. Depending on which wavelength is chosen, the contrast and maximum transmission level of the output function can be adjusted. Figure 2d shows the activation at 1553.4 nm (i.e. at the dashed line in Figure 2c), representing the operation with the highest contrast of 9 dB between the output states. It can clearly be seen that only above a threshold energy of 60% of the maximum pulse energy is a significant output generated. This non-linear response resembling the rectified linear unit (ReLU) function is crucial for neural activation functions, since it projects complex input data to higher dimensions enabling linear separation by the output neurons[30]. Note that the maximum pulse energy used to switch the PCM-cell in this experiment was 710 pJ employing optical pulses of 200 ns width. For setting the weights similar energies are used. In the experiments we operate our neurons with relatively long optical pulses in order to implement two-pulse switching[29] which relies on overlapping pulses in the time-domain. This approach is required for unsupervised learning as described below. We note, however, that lower switching energies can be employed by moving towards ps pulses[29,31]. It is also important to note that this operation scheme does not require continuous energy input for maintaining the state of the PCM weights due to their non-volatile response. Therefore, the energy budget per neuron to perform one operation is given by the switching energy for the ring resonator plus the energy required to reset the PCM element to its original state.

Having found the working point for our all-optical spiking neuron, supervised learning tests are now carried out. In this case, the synaptic weights of the network are set by an external supervisor (as for example done in software-based ANNs using the backpropagation algorithm). Here, a training set of data consisting of pairs of input patterns and the expected output is shown to the network. Depending on the deviation between the expected and actual output, the synaptic weights are adjusted in an optimization process until the solution is best approximated and the network is trained.

The experimental neural network used in our work is composed of two single neurosynaptic systems (of the type shown in Figure 1d) each consisting of four input (pre-synaptic) neurons connected to one output (post synaptic) neuron by four PCM synapses each. The weights of the first neuron were set to the pattern "1010" meaning that the first and third PCM synapse were in a high transmission state (contributing significantly to the activation energy) and the second and fourth were in a low transmission state (contributing less to the activation energy). The second neuron was trained in the same way to the pattern "1100". In Figure 3a and b the post-synaptic neuron output is plotted as a function of the input pattern. It can clearly be seen that in both cases (i.e. for both input patterns) the neurons were trained successfully and, based on the neuron's output, it can be easily concluded which pattern was presented to the network. Using only two output neurons on the same set of input neurons, our all-optical neuromorphic system can already solve simple image recognition tasks. By increasing the number of inputs per neuron and the number of neurons, more complex images and more difficult tasks, such as digit recognition, can be solved using this same basic approach, as we show in simulations given in the supplementary material.

Above we illustrated an example of supervised learning. This learning technique is feasible for many tasks but has the limitation that a training set with tuples of input patterns and expected outputs must be present. If the output is unknown, for example if a certain repeating pattern must be found from a data stream, then supervised learning is not applicable and unsupervised learning procedures are necessary. In an unsupervised approach, the network updates its weights on its own and in this way adapts to a certain pattern over time, without the need for an external supervisor.

In order to do this, an update rule needs to be defined. A common concept in unsupervised learning is spike-timing-dependent plasticity (STDP) following Hebb's postulate[32]. Here the change in the synaptic weight after an output spike depends on the relative timing between the input and the output spike of a neuron (i.e. the timing difference between pre- and post-synaptic neuron firings). If an input signal arrives right before an output spike was generated, that input signal is likely to have contributed to reaching the firing threshold and the corresponding weight will be increased. If the input pulse arrives after the output spike occurred, the synaptic weight will be decreased. The amount of potentiation (weight increase) or depression (weight decrease) is a function of the time difference between input and output spike, as described by Bi and Poo[33].

A similar but simplified learning rule is applied in our all-optical neuron approach. As the timing between incoming pulses and output pules in our case is fixed (because we operate the neuron in a clocked way, one complete input pattern per time step – see Methods section), there is no varying time delay between input and output events. We therefore adopt a simplified learning rule, increasing the synaptic weights of all inputs that contributed to a spike generation, and decrease the weights of all that did not. Experimentally we obtain this behaviour by overlapping (in time) the output pulses with the input pulses. Via a feedback waveguide, the neuron's output spike is guided back to the synaptic PCM-cells (see Figure 3c). If the input pulses are now long enough in time such that they overlap with the feedback pulse at the waveguide crossing where the synaptic PCM-cell is located, then the overlapping pulses have enough energy to switch the synapse into its low-absorbing amorphous state (weights $w_2$ and $w_3$ in Figure 3c). A feedback pulse that encounters a synapse without an input pulse will partly crystallize the corresponding PCM-cell because of the lower pulse energy, and therefore decrease the weight (weights $w_1$ and $w_4$). Due to the properties of the phase-change material used, the PCM-cells can only be amorphized in a single step[31], meaning that if the neuron fires, all contributing inputs will always be potentiated completely. Opposite to that, full crystallisation can be achieved in several steps[29] and the weights can correspondingly be decreased stepwise. Successful unsupervised learning can be accomplished using such weight update rules, as we show experimentally below for small-scale systems, and in the supplementary information (section 10) by simulation for larger-scale systems.

Figures 3d and e show the development over time of the four synaptic weights during unsupervised by a single neuron. Initially all the PCM-synapses are in the amorphous (high transmittance) state. When the input pattern '0110' is repeated, the neuron adapts to it over time, until the neuron has finally learned this pattern without any intervention from an external supervisor. The neuron is now specialized to recognize this particular pattern. From Figure 3d it is clear that the weights $w_2$ and $w_3$, corresponding to the inputs three and four, stay almost constant over time, as the overlapping input and feedback pulses preserve their amorphous state. In contrast, weights $w_1$ and $w_4$ are depressed stepwise with each epoch.

Having successfully demonstrated a single-neuron neurosynpatic system as a fundamental building block for photonic neural networks, a way of connecting these artificial neurons into larger networks is now developed. An architecture exploiting individually addressable, interlinked, photonic layers is thus implemented, as shown schematically in Figure 4a. The whole network consists of an input and an output layer which are optically connected via N hidden layers. Each hidden layer takes the output of the previous layer as an input and passes its outputs to the next layer. The input layer is the optical interface to the real world, taking the data to be processed and distributing it to the next level in the network.

A single layer of the network consists of a collector, a distributor and its neurosynaptic elements. The collector gathers all the outputs from the previous layer, which are than equally distributed to the N neurons within the layer (fully connected network) by the distributor. The photonic neurons themselves operate as described in detail before: a phase-change synapse weights the inputs and a WDM multiplexer builds the sum, which is passed to the activation unit that decides if a neuronal output pulse is transmitted. In this architecture, each layer is addressed optically with its own waveguide for generating the probe signal. Therefore, the optical power in the layer is not limited by the transmitted optical response from a previous layer.

Figure 4b describes how the constituent parts of a layer translate into the actual photonic circuit. The outputs from a previous layer are multiplexed onto a single waveguide using ring resonators (thus building the collector). This signal is then equally distributed to the neurons within this layer, again using ring resonators for demultiplexing (thus building the distributor). By choosing the gap between the feeding waveguide and ring resonator, the coupling efficiency can be tuned (see supplementary figure S8.) Following the formula for the coupling efficiency $c\_eff,i=1/(N'+1-i)$ with N' neurons on the layer and the neuron position i. For example, in a layer with four neurons this means, that $1/(4+1-1)=1/4$ of the light is coupled to the first neuron, 1/3 of the remaining light is passed to the second neuron, 1/2 to the third and the residual light to the third neuron. The circuit-diagram of the actual photonic neuron, the neuro-synaptic system, was shown in Figure 1c and is the same as used in the experiments described above. The output pulses of a layer can then be connected to the collector of the next layer.

We note that using the above approach no waveguide crossings are needed for distributing the signal to the neurons, thus preventing crosstalk and losses. Because the output pulses are generated for each layer individually, there is also no accumulation of errors and signal contamination over subsequent layers. This fact also simplifies the timing of the network as each layer can be processed step by step: First, the input pulses are sent, and the activation units are switched where appropriate. Second, the output probe pulses are sent and transmitted to the next layer (if the threshold for switching was reached). In a final step, the PCM-cells on the rings have to be reset to their initial state.

Figure 5 shows the experimental implementation of a full layer of the proposed neural network design consisting of four neurons with 15 synapses each. The full device is composed of more than 140 optical elements, optical micrographs of the photonic circuit are presented in section 2 of the supplementary materials. This network is capable, by way of example, to differentiate between four 15-pixel images, here representing the four letters A, B, C and D. In this system, the neurons are optically fed via an integrated WDM distributor with 15 ring resonators per neuron, while the collector is implemented off-chip using fiber-based WDM components. The images shown in Figure 5b (corresponding to the letters A-D) are encoded in optical pulse patterns which are presented to the on-chip network. Each pixel corresponds to the resonance wavelength of one of the ring resonators within a neuron, as indicated by the numbers superimposed on the pixels in fig 5b. These wavelengths are aligned to WDM channels 27-41 in the telecommunication C-band. In the experiment we present the "white" pixels to the network such that the pulse pattern corresponding to an 'A' is, for example, represented by an optical pulse consisting of wavelengths 1, 3, 5, 8 and 14. These wavelengths are multiplexed onto the input waveguide as described in the supplementary materials and equally split to the synapses of the four neurons by the distributor. After adjusting the synapses (PCM-cells) corresponding to the patterns 'A', 'B', 'C' and 'D' using optical pulses, as described previously for the single neuron before (Figure 3), the four different pulse patterns are sent to the input of the device. Subsequently, the change in the output spike intensity is observed for all four neurons as shown in Figure5b. As desired, all four output neurons are only activated when the learned pattern is shown: Neuron 1 only fires if pattern 'A' is shown, neuron 2 only reacts to pattern 'B' and so forth. Thus, the network is able to successfully classify the four 15-pixel images.

Using the architecture outlined above we further simulate the performance of a scaled-up, multilayer version of the network of fig. 5a for carrying out a much more complex task of language identification. The network for this task consists, as shown,in Figure S16a),b) in the supplementary materials of four input neurons, three hidden layer neurons and two output neurons. The network is assembled using the scaling architecture described in Figure 4 and built up using model representations of photonic neurons according to the measured experimental data (activation function shown in Figure 2d)). This particular network is then used to detect if the language of a given input text is either English or German (the sample texts are taken from[34]). In a first step the ratio of each vowel ("a", "e", "i", "o", "u") and the total number of characters in the input text is calculated (preprocessing). In a second step the five obtained ratios are fed to the inputs of the neural network and the outputs are computed. Each neuron in the simulated network uses the measured optical response from the on-chip neurons (see supplementary materials section 10). Already with a count of about 35 words in the input text, an accuracy above 90% for the language detection is attained. With 150 words the accuracy reaches 99.6%.

Integrated phase-change photonic networks, designed and implemented as described above, are capable of simple pattern recognition tasks and can adapt to specific patterns. When operated with a waveguide feedback loop, they are capable of learning without an operator needed, and can do this in a non-volatile fashion using phase-change materials. The large contrast in absorption of light between their amorphous and crystalline state of matter makes phase-change materials an attractive and simple solution to be integrated as synaptic weighting mechanism. Compared to conventional computers that can only simulate the parallelism of neural networks, our all-optical neurons are intrinsically suited for mimicking biological neural networks. Compared to speeds of biological neural networks (~ milliseconds) our proposed neurons could operate several orders of magnitude faster, giving rise to substantial potential in dealing with large amounts of data in a short amount of time. Working exclusively in the optical domain, the spiking neurosynaptic network benefits from high-bandwidth and fast data transfer rates intrinsic to light. Moreover, using a layered circuit architecture, we present a pathway to scaling our network to more complex systems which could be realized with foundry processing. This way also the off-chip components (used here for experimental expediency only) such as laser sources, optical amplifiers and modulators, can be integrated into a full system. Our integrated and novel design combines, via wavelength division multiplexing techniques, the outputs of multiple phase-change synapses to excite layered spiking phase-change neurons and holds promise for realizing all-optical neural networks capable of addressing the upcoming challenges of big data and deep learning.

## Methods

**Device fabrication**

The nanophotonic circuits used in this work are realized using electron-beam lithography (EBL) with a 100-kV system (Raith EBPG 5150). In a first step, opening windows for lift-off processing of alignment markers made from gold are exposed in the positive tone resist Polymethylmethacrylat (PMMA) on a silicon wafer (Rogue Valley Microdevices) with a 3300 nm silicon oxide and 344 nm silicon nitride layer on top. After development in 1:3 MIBK:Isopropanol for 2 min, a stack of 5 nm chromium, 120 nm gold and 5 nm chromium again are evaporated via electron-beam physical vapour deposition (PVD). The lift-off step to remove the PMMA is performed in acetone, leaving the gold markers for the alignment in the next EBL-steps.

In the second lithography step the photonic structures are defined. TI Prime is used as an adhesion agent for the negative-tone ebeam resist maN 2403. The photonic circuitry is developed in MF-319 for 60 s and afterwards placed on a hotplate at 105°C for two minutes of reflow processing to reduce surface roughness. By reactive ion etching in a CHF3/O2 plasma, the resist mask is transferred into the sample till the silicon nitride is fully etched. The remaining resist on the structures is removed in an oxygen plasma for 10 minutes.

The last EBL step consists of writing windows for the deposition of the phase-change material and is executed in the same way as defining the marker windows. After development, 10 nm of the phase-change material GST are sputter-deposited and covered by a 10 nm film of indium tin oxide (ITO) to prevent oxidation of the GST. The GST and ITO capping layers were deposited using RF sputtering with an argon plasma (5 mtorr working pressure, 15 sccm Ar, 30 W RF power, and base pressure of $2\times10$-6 Torr). Finally, the GST is crystallized on a hot plate for about 10 minutes at 210°C. The photonic circuits are composed of single mode waveguides at 1550 nm with a width of 1.2 µm.

**Measurement setup**

The experimental setup used to operate the all-optical neurons comprises pattern generation and read-out of the individual weights, as sketched in the Supplementary materials. The optical read-out is achieved via transmission measurement using a continuous wave laser (Santec, TSL 510) and four low-noise photodetectors D1-D4 (New Focus, Model 2011) that are monitored on a computer. In order to couple light efficiently onto the chip, an optical fiber array is aligned with respect to the on-chip grating couplers to provide multi-port input. For optimal coupling efficiency to the chip the polarization is optimized with a set of polarization controllers.

Pattern generation as input for the on-chip neuron is accomplished with four cw-lasers set to different wavelength matching the on-chip multiplexer. Off-chip the four light paths are combined using a fiber multiplexer and desired optical pulses are created using an electro-optical modulator (EOM) and a computer controlled electrical pulse generator (Agilend, HP 8131A). After amplifying the pulses with an erbium-doped fiber amplifier (Pritel, EDFA) the pulses are de-multiplexed again and guided to the on-chip device. Using circulators pump and probe light are counter-propagating through the device enabling efficient separation of the beam paths. Arbitrary input patterns are then selected by switching on and off the shutters of the pump lasers.

Similar to the input pulses, the output pulses are also created from a cw-laser in combination with an EOM and amplified by an EDFA. A small portion of the output is measured with detector D0, while the remaining light is amplified and send to the ports F1-F4 as a feedback pulse for weight adjustments. Turning off the feedback amplifier puts the device in a supervised learning mode. The setup used for the four neuron-network is shown in supplementary figure S2.

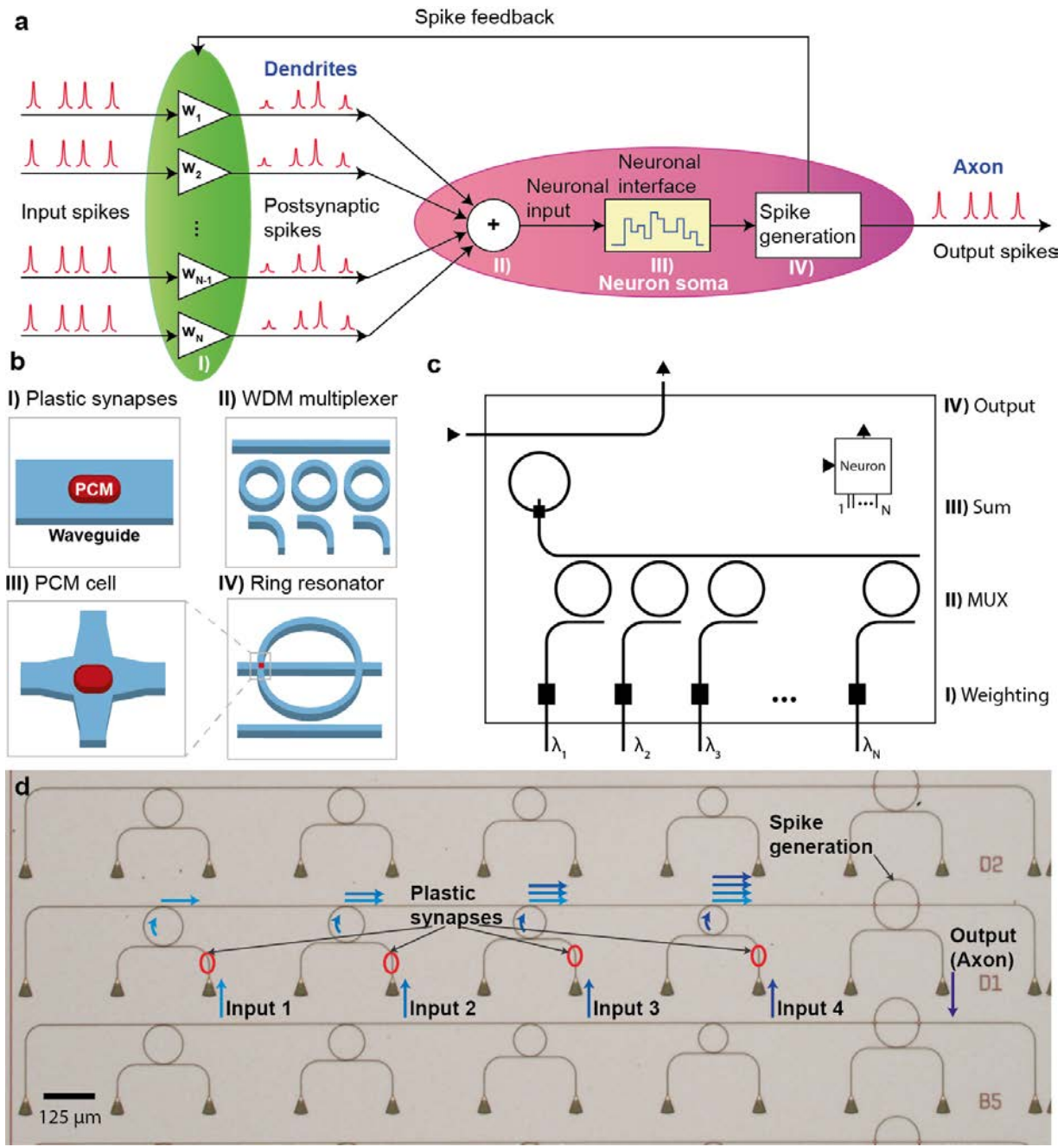

**Figure 1. All-optical spiking neuronal circuits. a-b)** Schematic of the network realized in this work consisting of several pre-synaptic input neurons and one post-synaptic output neuron connected via PCM-synapses. The input spikes are weighted using PCM-cells and summed up using a WDM multiplexer. If the integrated power of the postsynaptic spikes surpasses a certain threshold, the PCM-cell on the ring resonator switches and an output pulse (neuronal spike) is generated. **c)** Photonic circuit diagram of an integrated optical neuron with symbol block shown in the inset (top right). Several of these blocks can be connected to larger networks using the wavelengths inputs and outputs as described in more detail in Figure 5 **d)** Optical micrograph of three fabricated neurons (B5, D1 and D2) showing four input ports. The four small ring

resonators on the left are used to couple light of different wavelengths from the inputs to a single waveguide, which then leads to the phase-change material cell at the crossing point with the large ring. The triangular structures on the bottom are grating couplers used to couple light onto and off the chip.

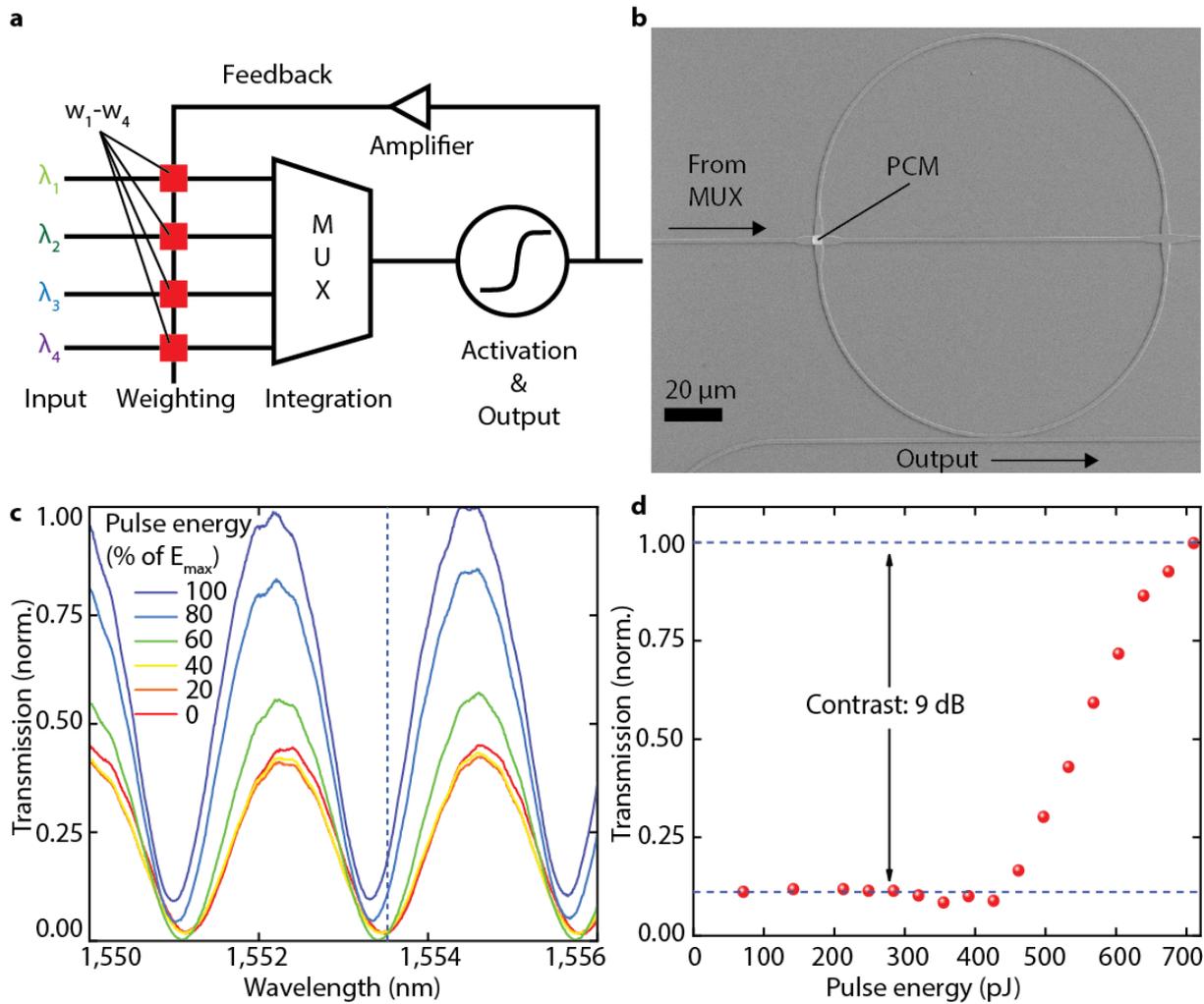

**Figure 2. Spike generation and operation of the artificial neuron. a)** Schematic of the photonic implementation of a phase-change neuron circuit. Light of different wavelength is weighted by phase-change elements $w_1$-$w_4$ and summed up by a multiplexer to a single waveguide. If this activation energy surpasses a threshold, an output pulse is generated, and the weights are updated. **b)** Scanning electron micrograph of a ring resonator used to implement the activation function. By switching the PCM-cell on top of the waveguide crossing, the resonance condition of the resonator can be tuned. The waveguide on the bottom of the ring is used to probe the resonance and generate an output pulse. **c)** Transmission measurement of the device in **b)** and its dependence of pulse energy. The resonance shifts towards shorter wavelength with increasing pulse energy send to the PCM-cell on the ring. At the same time the transmission increases because of reduced absorption in the PCM-cell and thus changes the coupling between ring and waveguide. **d)** Normalized transmission to the output at a fixed wavelength (dashed line in **c)**) showing the activation function used to define the firing threshold of the neuron.

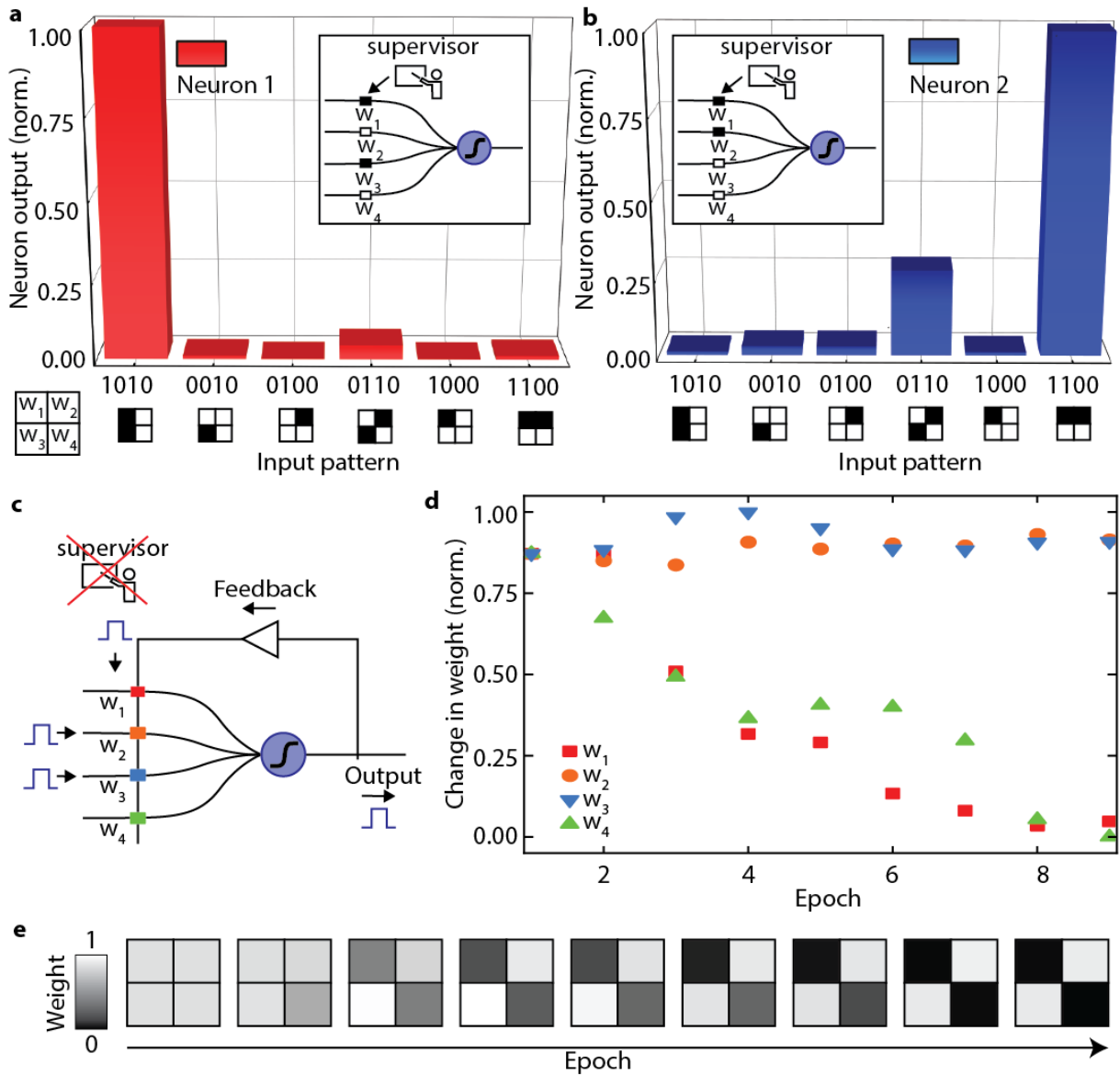

**Figure 3. Supervised and unsupervised learning with phase-change all-optical neurons. a)** and **b)** show the neuron output of two individual neurons when presented with different input patterns. Neuron one learned to recognize pattern '1010', while neuron two generates an output signal when '1100' is shown. In this example the eight weights of the neural network were set by an external supervisor. **c)** Schematic illustrating the unsupervised learning mechanism in an all-optical neuron. If an output spike is generated, the synaptic-weights where input and feedback pulses overlap in time are potentiated, while the weights that are only hit by the single feedback pulse are depressed. **d)** Change of the four synaptic weights over time when the pattern '0110' is repeatedly shown starting from fully amorphous (high transmitting) weights. The weights where input- and feedback pulse overlap stay almost constant over several epochs. The other weights where only the feedback pulse is shown decrease continuously. **e)** Development of the weights over time, clarifying that the information of the pattern is encoded in the weights.

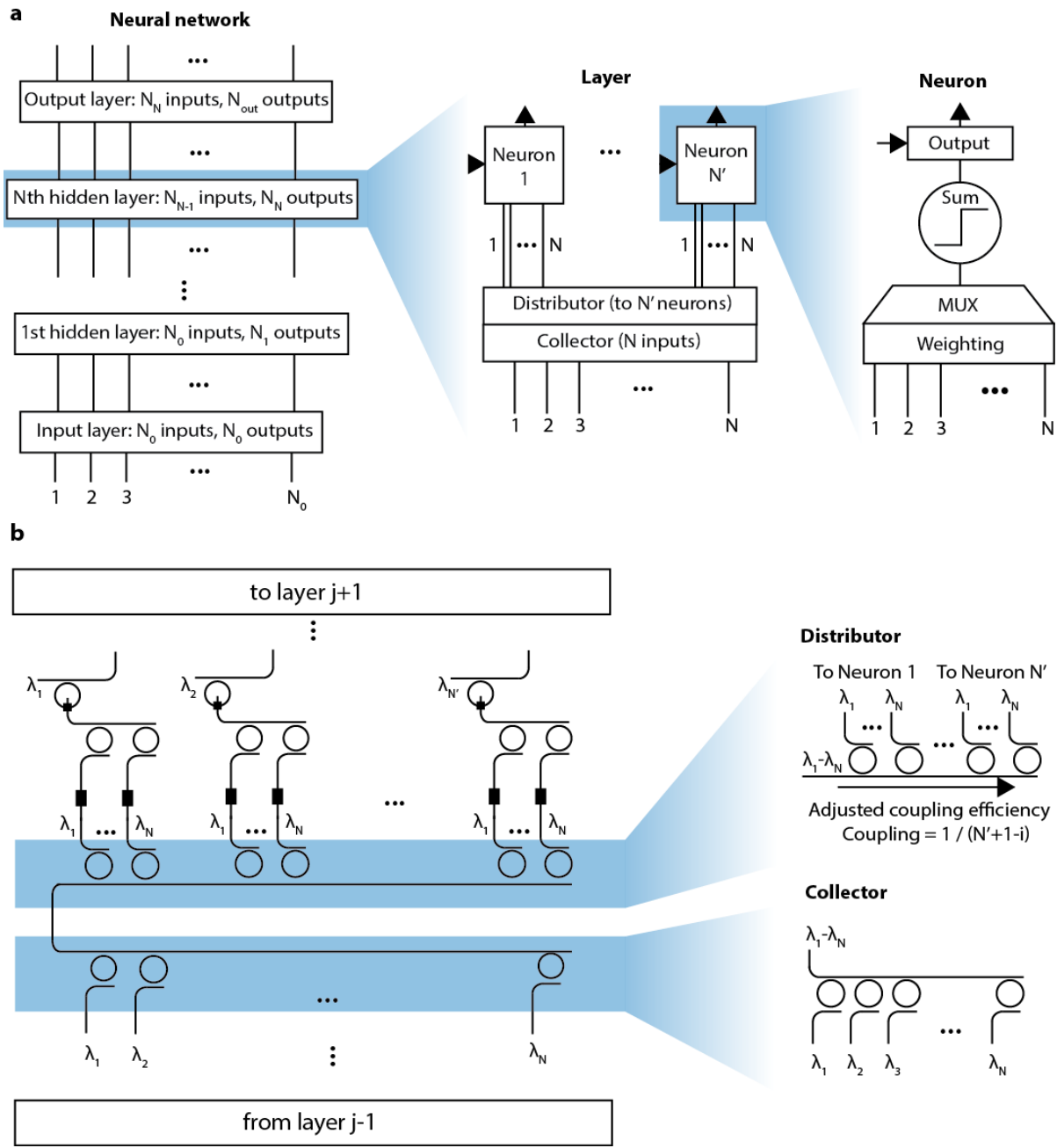

**Figure 4. Scaling architecture for all-optical neural networks. a)** The general neural network is composed of an input layer, an output layer and several hidden layers. Each of these layers consist of a collector gathering the information from the previous layer, a distributor that equally splits the signal to individual neurons and the neuronal and synaptic elements of the layer itself. Each neuron has a weighting unit and a multiplexer to calculate the weighted sum of the inputs. The sum is then fed to an activation unit which decides if an output pulse is generated. **b)** Photonic implementation of a single layer from the network. The collector unites the optical pulses from the previous layer using a WDM multiplexer. A distributor made from the same rings as the collector but with adjusted coupling efficiency equally distributes the input signal to the PCM synapses of each neuron.

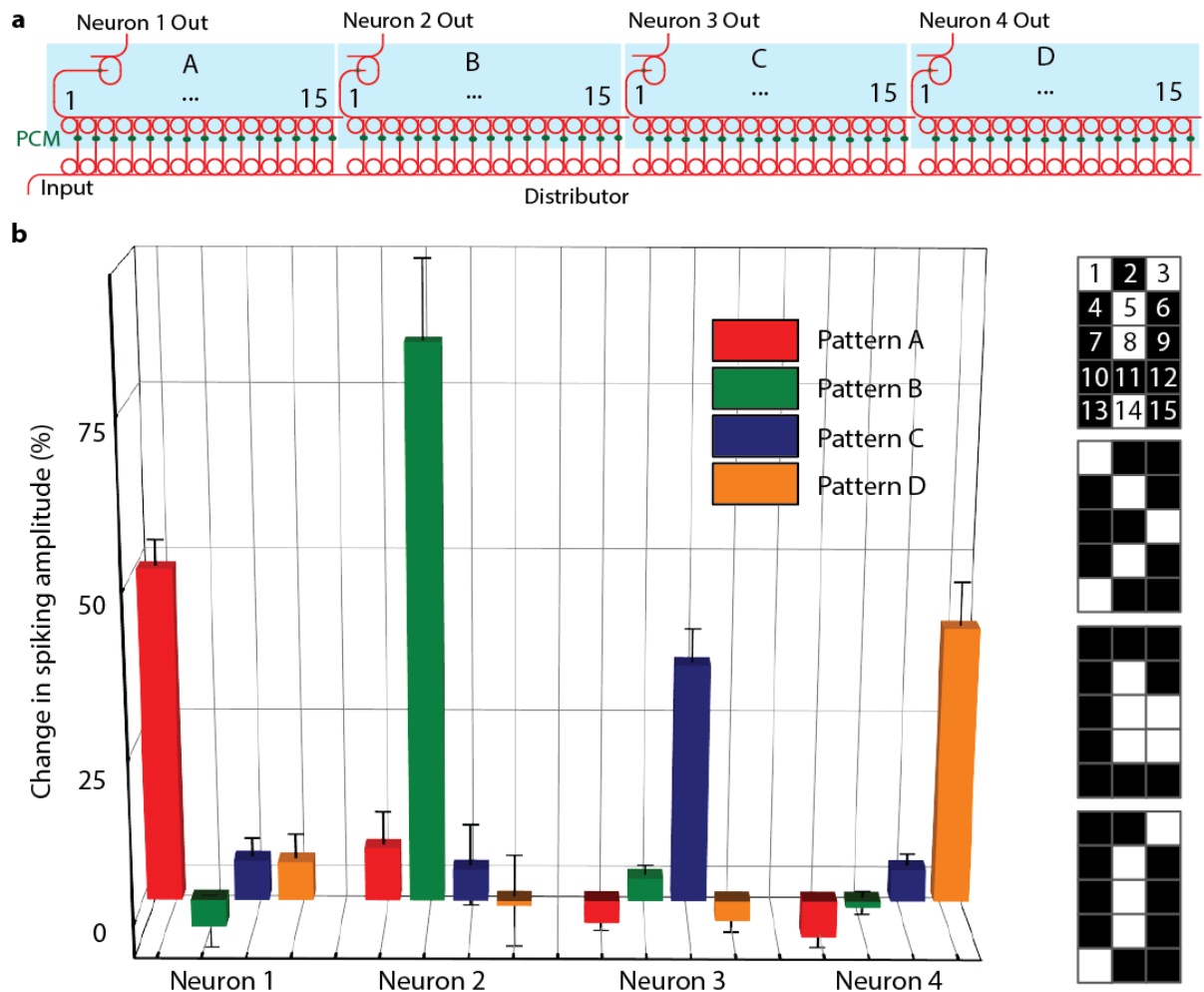

**Figure 5 Experimental realisation of a single layer spiking neural network. a)** The device consists of four photonic neurons, each with 15 synapses. Each synapse corresponds to a pixel in a 3x5 image (see b)) and is encoded in the wavelengths corresponding to the ring multiplexers (see numbering in b)). The full device comprises an integrated photonic circuit built up from 140 optical components. **b)** The change in spiking amplitude is shown for the four trained patterns illustrated on the right hand side. The neural network successfully recognizes the four patterns as each neuron only responds (spikes) to one of the patterns.